\documentclass[a4paper,11pt]{article}
\pdfoutput=1 % if your are submitting a pdflatex (i.e. if you have
\usepackage{jinstpub} % for details on the use of the package, please
\usepackage{subfigure}

\title{\boldmath Numerical study on the effect of design parameters and spacers on RPC signal and timing properties}

\author[a,c,1]{A. Jash,\note{Corresponding author.}}
\author[a]{N. Majumdar,}
\author[a]{S. Mukhopadhyay,}
\author[a]{S. Saha}
\author[b]{and S. Chattopadhyay}

\affiliation[a]{Applied Nuclear Physics Division, Saha Institute of Nuclear Physics, Kolkata, India}
\affiliation[b]{Experimental High Energy Physics Division, Variable Energy Cyclotron Centre, Kolkata, India}
\affiliation[c]{Experimental High Energy Physics Division, Homi Bhabha National Institute, Mumbai, India}

\emailAdd{abhik.jash@saha.ac.in}

\abstract{% Abstract for the document "rpc paper for arXiv".
Numerical calculations have been performed to understand the reason for the observed non-uniform response 
of a Resistive Plate Chamber (RPC) in a few critical regions such as near edge spacers and corners of 
the device. In this context, the signal from a RPC due to the passage of muons through different regions has been 
computed. Also, a simulation of RPC timing properties is presented along with the
effect of the applied field, gas mixture and geometrical components.
  }

\keywords{Resistive-plate chambers; Detector modelling and simulations II (electric fields, charge 
transport, multiplication and induction, pulse formation, electron emission, etc)}

\arxivnumber{1605.02154} % only if you have one

\proceeding{13$^{\text{th}}$ WORKSHOP ON RESISTIVE PLATE CHAMBERS AND RELATED DETECTORS,\\
22$-$26 FEBRUARY 2016,\\
GHENT UNIVERSITY, BELGIUM\\
}

\begin{document}
\maketitle
\flushbottom

  % Introduction for the document "rpc paper for arXiv".
\section{Introduction}
The India-based Neutrino Observatory (INO) \cite{INO}, the proposed underground laboratory 
facility in India will host a magnetized Iron CALorimeter (ICAL) to conduct precise measurements 
of neutrino oscillation related parameters by studying atmospheric neutrinos. A stack of 
Resistive Plate Chambers (RPC) in the ICAL detector will provide the timing and 3D spatial 
information of passing muons, produced in the interaction of the atmospheric neutrinos with the 
iron layers. About 30,000 RPCs of dimension 2 m $\times$ 2 m will be deployed in the ICAL setup. 
In the last few years, a good amount of R\&D effort has been delivered on various issues related to 
the performance optimization of ICAL. An observation of reduced response towards the edges and 
corners of the RPC \cite{RPC_stack_manas} has been regarded as a major motivation for 
the present work.
\\
Numerical simulation is a good tool to investigate the different processes behind the operation 
of the detector to analyze the experimental results. This can also be used to optimize different 
design parameters to improve the detector performance. In the present study, the signal induced 
on the readout strips for the passage of muons through the RPC has been simulated. To study the effect 
of the device geometry, the signal has been computed at different regions of the device for different 
voltages. Some of the results have been compared to actual measurements to validate the calculation
procedure. An appropriate approach for this study would be to find out the signal in the presence of a 
dynamic electric field  as the space charges created in the avalanche process tend to modify it \cite{paper_Lippmann}.
Also the time dependence of the electric field owing to the finite bulk resistivity of the RPC plate 
should be taken into account in order to carry out an extensive signal simulation. However, the 
present work involves calculations assuming a static electric field configuration
where the RPC is described as a multi-dielectric planar capacitor. 
\\
Accurate timing information from the RPC layers of the ICAL setup will help identifying the direction of 
the passing muon tracks. The parameters of importance in the timing performance of RPC are the 
average signal arrival time and the time resolution, which in turn rely upon the electric field configuration 
of the device and the gas mixture used.
In this work, the variation of timing properties of a RPC with the applied voltage has been studied. 
In order to study the effects of the device, the simulation has been carried out generating 
the timing response for several critical regions. The dependence on the gas mixture has been 
investigated through simulations with different SF$_{6}$ concentrations. This component of the 
gas mixture is known to be important for restricting the streamer mode activity. The results have 
been compared to measurements wherever available.
\\
The experimental setup and a brief description of the numerical methods are presented 
in section \ref{sec:experiment_simulation}.
Section \ref{sec:Result} is dedicated to the discussion of the results and the final remarks 
are made in section \ref{sec:conclusion}.

  \section{Experiment and Simulation Methodologies}
\label{sec:experiment_simulation}
\begin{figure}[ht]
 \centering
  \includegraphics[width=.8\textwidth]{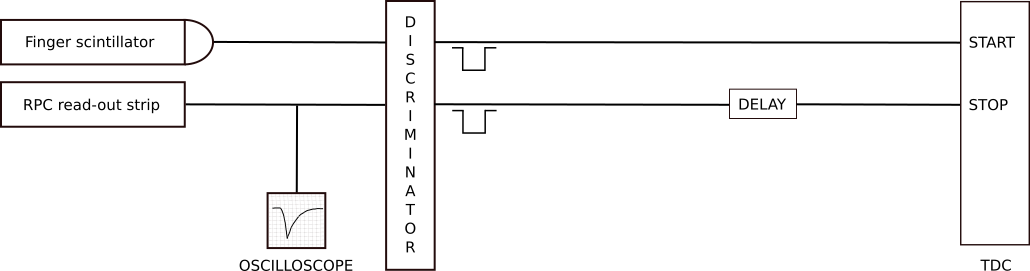}
 \caption{Schematic diagram of the experimental setup for RPC signal and timing measurements.}
\label{fig:expt_schematic}
\end{figure}

One bakelite RPC of dimension 30 cm $\times$ 30 cm with 2 mm gas gap has been operated with 
a gas mixture of R-134A and Isobutane in 95 : 5 ratio. The schematic diagram of the experimental 
setup used for carrying out the measurements of the RPC and timing properties is
shown in figure \ref{fig:expt_schematic}. A finger scintillator of width 4 cm has been placed aligned 
with one of the RPC readout strips. 
\\
In our numerical simulation, a bakelite RPC of dimension 30 cm $\times$ 30 cm has been modelled as 
described in \cite{paper-RPC_field}. The Garfield \cite{web_Garfield} simulation framework has been 
used to calculate the signal and timing properties of RPC along with its toolkits like neBEM \cite{neBEM}, 
HEED \cite{HEED} and Magboltz \cite{Magboltz} performing different stages of the calculation. 
neBEM is used to calculate the electrostatic field map, while HEED computes the 
primary ionization produced by the relativistic charged particles in the gas mixture. The transport 
properties of electrons in the gas mixture are generated by Magboltz and finally Garfield simulates 
the drift of primary and avalanche electrons to produce the current induced on readout strips due 
to the movement of this charge cloud. The field in the gas chamber has been produced by applying 
bias voltage across the readout panels, to overcome some technical restrictions discussed 
in \cite{paper-RPC_field}.
\\
The timing properties of the RPC have been computed by passing 2 GeV muons with incidence 
angle varying randomly in the range 0$^{\circ}$ - 10$^{\circ}$ through a region away from any 
imperfection (regular region). The passage of a muon track triggers an avalanche of electrons and 
ions which move towards respective electrodes and induce currents on the readout strips. The current 
signals for the passage of 5000 muons have been calculated using Garfield. 
The average signal arrival time and the time resolution were determined as the mean and RMS, 
respectively, of the distribution of the time corresponding to the crossing of 20\% of the signal amplitude.
The same calculations have been performed setting different threshold values in the range  10\% - 50\% 
of the signal amplitude.  No significant dependence on the threshold value has been observed 
except for a shift in the average signal arrival time towards higher values for higher thresholds, which 
is obvious from the fact that higher current values appear at a later time. Small fluctuations in the 
currents have been observed at very small values of time which has been avoided by using an 
optimum threshold set to 20\% of signal amplitude.
In the present study, no effect of ion movements or electronics has been considered.

  \section{Result}
\label{sec:Result}
\subsection{RPC Signal}
\label{result_signal}
A typical signal waveform as seen on an oscilloscope, due to the passage of muons through the 
\begin{figure}[ht]
 \centering
  \subfigure[]{
  \includegraphics[width=.35\textwidth, height =5cm]{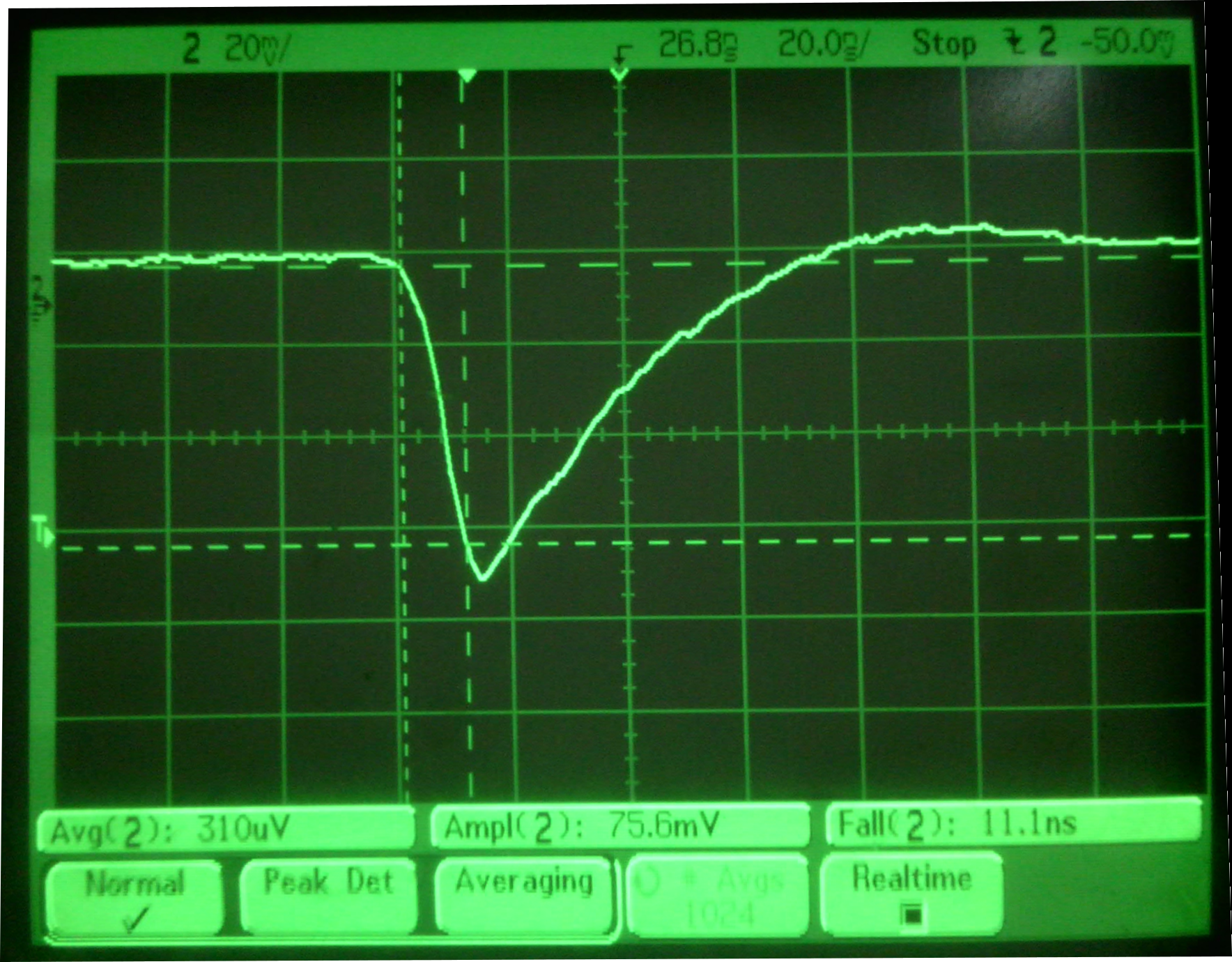}
  \label{fig:signal_experimental}
  }
  \hspace{0.5cm}
  \subfigure[]{
  \includegraphics[width=.35\textwidth, height =5cm]{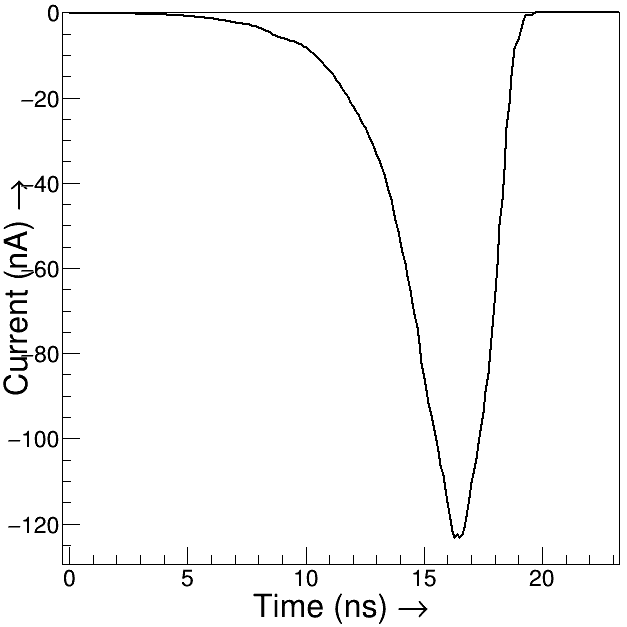}
  \label{fig:signal_simulated}
  }
 \caption{\subref{fig:signal_experimental} Typical RPC signal on an oscilloscope, 
 \subref{fig:signal_simulated} Simulated signal waveform (average of 50 events).}
\label{fig:signal_shapeComparison}
\end{figure}
RPC is shown in figure \ref{fig:signal_experimental}.  Figure \ref{fig:signal_simulated} shows 
a typical signal waveform as calculated numerically for the passage of 2 GeV muons at an 
angle 5$^{\circ}$ with the vertical direction. 
\begin{figure}[h]
 \centering
  \subfigure[]{
  \includegraphics[width=.41\textwidth]{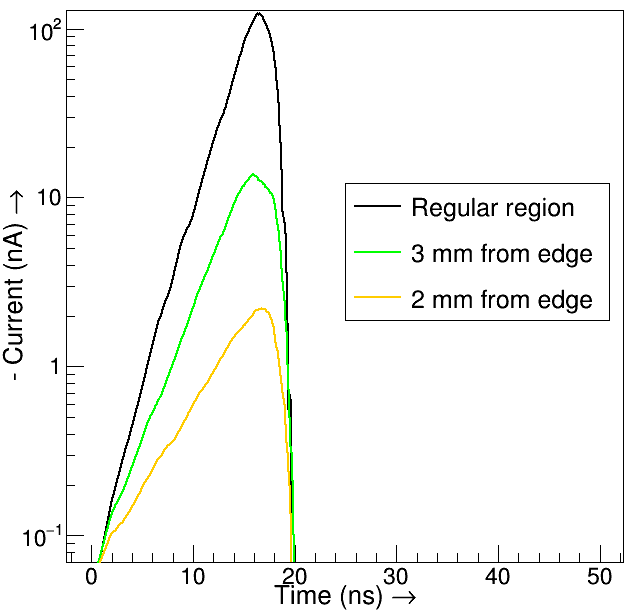}
  \label{fig:signal_12p1kV}
  }
  \hspace{0.2cm}
  \subfigure[]{
  \includegraphics[width=.41\textwidth]{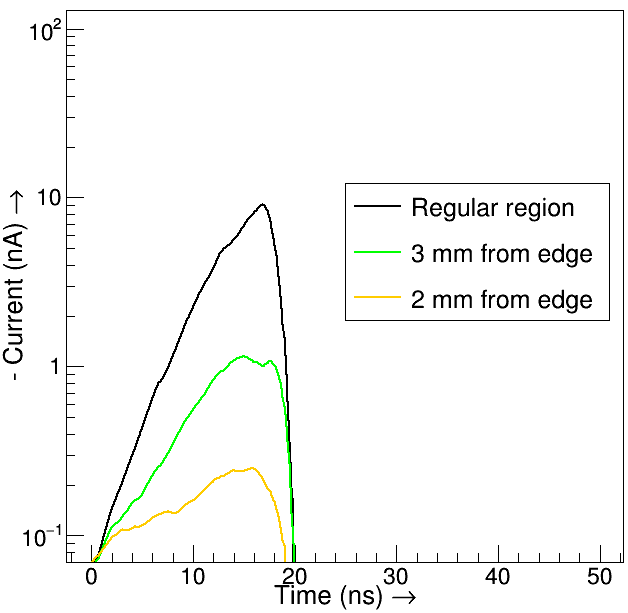}
  \label{fig:signal_11p7kV}
  }
 \caption{Average signal induced due to the passage of 100 muons of energy 2 GeV through different 
 regions of the RPC for the voltages \subref{fig:signal_12p1kV} 8.6232 kV and \subref{fig:signal_11p7kV} 
 8.3381 kV across the coats.}
\label{fig:signal_edgeEffect}
\end{figure}
As neither the effect of ion movement nor the electronics have been taken into account, only the rising 
edge of the signal of 7 ns has been compared with the experimental form with a rise time of 11 ns.
To see the effect of edge spacers, 100 muons with 2 GeV energy have been sent through a
regular region and regions near the edge. The waveform of signals averaged over the 100 events 
in those regions have been calculated and are shown in figure \ref{fig:signal_edgeEffect} for two 
different voltages.
It can be seen that the amplitude of the signal falls as one approaches the edge spacer which is obvious 
as the electric field suffers from an edge effect \cite{paper-RPC_field}. Consequently, the loss of 
response at those regions can lead to less efficient or dead regions near the edges of the RPC.
\subsection{RPC Timing Properties}
\label{result_timing}
\begin{figure}[ht]
 \centering
  \subfigure[]{
  \includegraphics[width=.45\textwidth]{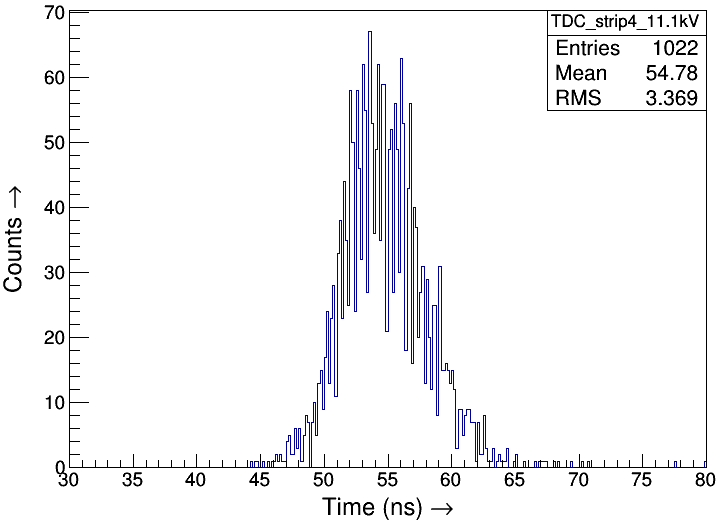}
  \label{fig:TDC_spectra_5p5kV}
  }
  \hspace{0.1cm}
  \subfigure[]{
  \includegraphics[width=.45\textwidth]{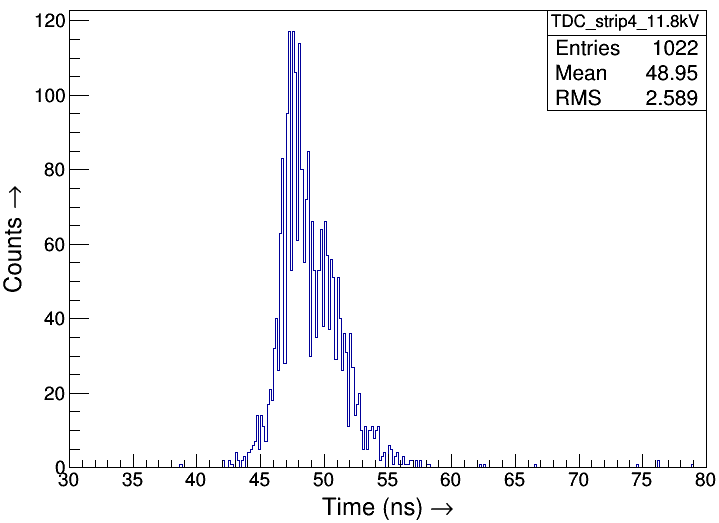}
  \label{fig:TDC_spectra_5p9kV}
  }
 \caption{Typical TDC spectra of RPC when the applied voltage is \subref{fig:TDC_spectra_5p5kV} 11.1 kV, 
 \subref{fig:TDC_spectra_5p9kV} 11.8 kV.}
\label{fig:TDC_spectra}
\end{figure}

\begin{figure}[ht]
\centering 
	\includegraphics[width=.95\textwidth]{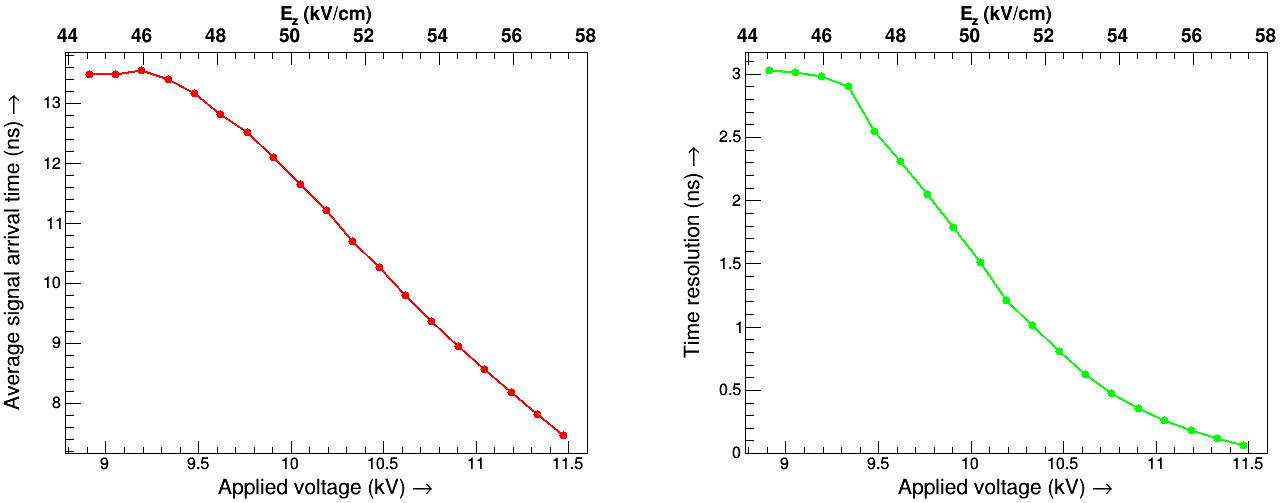}
	\caption{Variation of average signal arrival time and time resolution with the applied voltage 
	found numerically for R-134A : Isobutane = 95 : 5.}
	\label{fig:t_sigma_vs_HV}
\end{figure}

\begin{figure}[ht]
\centering
	\includegraphics[width=.95\textwidth]{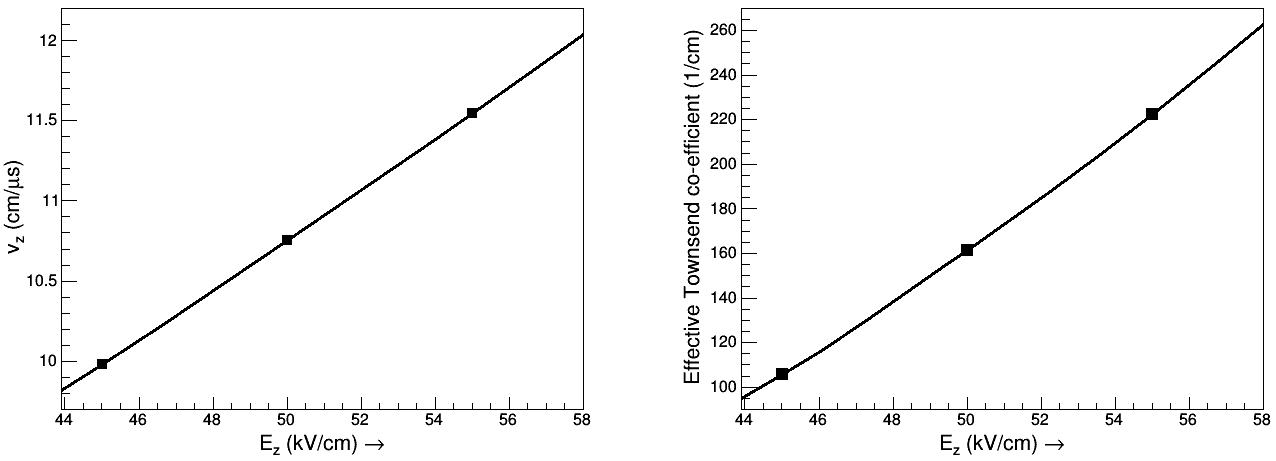}
	\caption{Variation of electron drift velocity (V$_{z}$) and effective Townsend co-efficient of 
	the gas mixture R-134A : Isobutane = 95 : 5 with E$_{z}$.}
	\label{fig:Vd_Dt_vs_Ez}
\end{figure}

\begin{figure}[h]
	\includegraphics[width=0.95\textwidth]{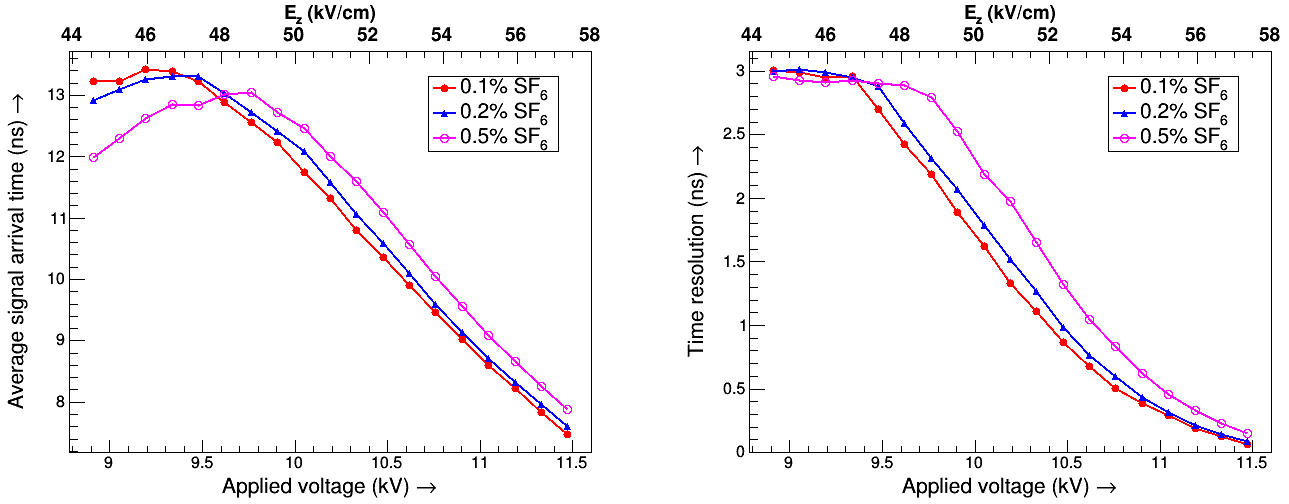}
	\caption{Variation of average signal arrival time and time resolution with the applied voltage for 
	different SF$_{6}$ percentages.}
	\label{fig:t_sigma_vs_HV_manyGas}
\end{figure}

The typical TDC spectra  measured for two different voltages are shown in figure \ref{fig:TDC_spectra}.
The mean  and the RMS values of the histogram give the average signal arrival time and  the time resolution
respectively. 
\begin{figure}[h]
\centering
\includegraphics[width=.95\textwidth]{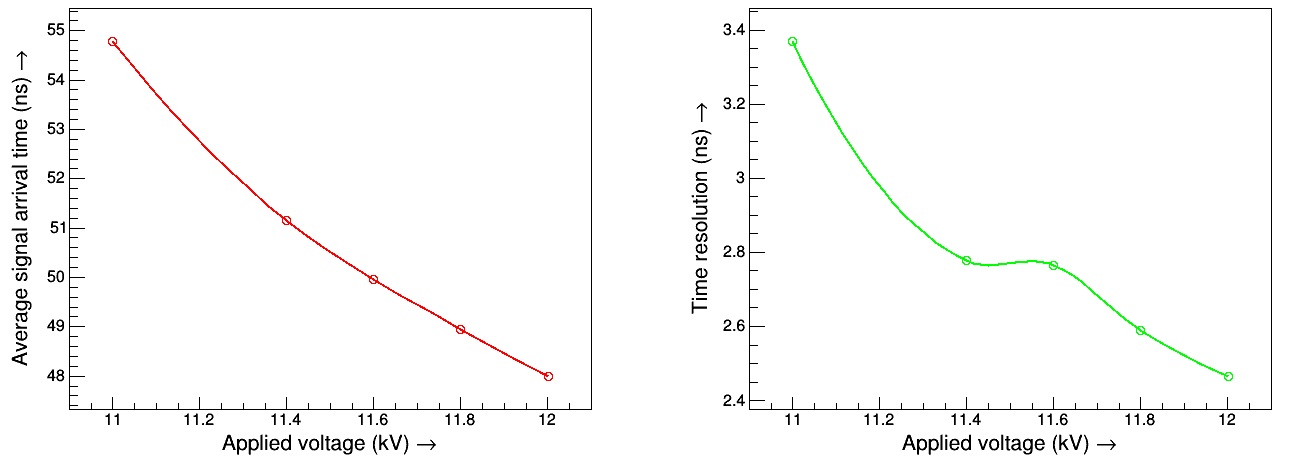}
\caption{\label{fig:mean_sigma_Vs_HV} Variation of average signal arrival time and RPC time resolution 
with applied voltage from experimental data.}
\end{figure}
The variation of average signal arrival time and time resolution with the applied voltage is shown in 
figure \ref{fig:mean_sigma_Vs_HV}. The numerical calculation of the same quantities is shown in 
figure \ref{fig:t_sigma_vs_HV}. 
Both the average signal arrival time and time resolution of the RPC decrease with increasing field, 
as the values of the drift velocity of electrons (V$_{z}$) and the effective Townsend co-efficient 
increase with the field, which follows from the relation of time resolution to these parameters discussed 
in \cite{paper-Riegler_timing}. The variation of these two gas parameters for the gas mixture 
R-134A : Isobutane = 95 : 5 with the increase in electrostatic field, as calculated using Magboltz 
 is shown in figure \ref{fig:Vd_Dt_vs_Ez}.
\\
To see the effect of different geometrical components on the timing properties, a simplified approach 
has been taken where their effect on the drift of electrons has been determined. Electrons have been 
released at different regions, all at a fixed distance (1.7 mm) from the bakelite plate on the side of 
anode and the time taken by them to reach the anode has been filled in a histogram. This calculation 
has been done keeping the applied voltage constant for which the value of E$_{z}$ at a regular 
position is 42.76 kV/cm.
\begin{figure}[ht]
 \centering
  \subfigure[]{
  \includegraphics[width=.3\textwidth]{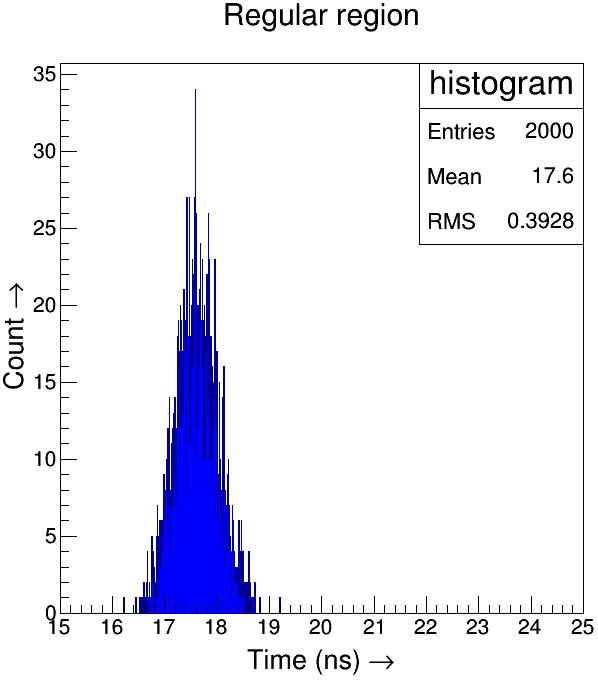}
  \label{fig:timingHisto_regular}
  }
  \hspace{0.1cm}
  \subfigure[]{
  \includegraphics[width=.3\textwidth]{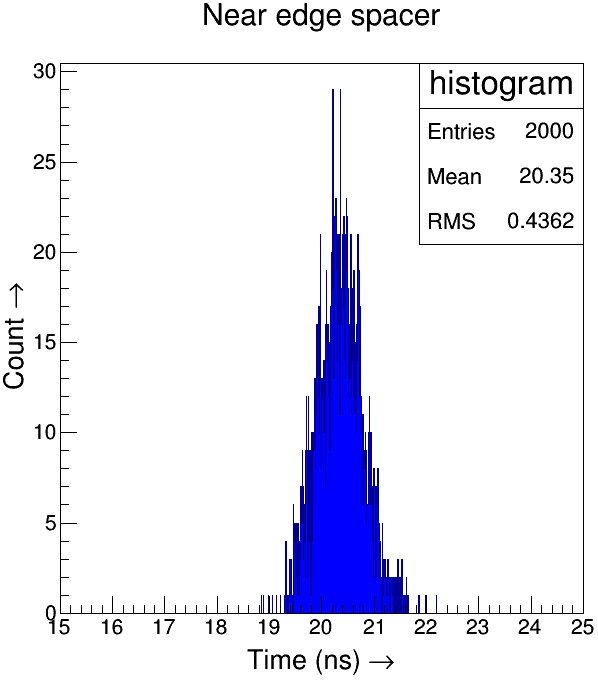}
  \label{fig:timingHisto_nearESpacer}
  }
  \hspace{0.1cm}
  \subfigure[]{
  \includegraphics[width=.3\textwidth]{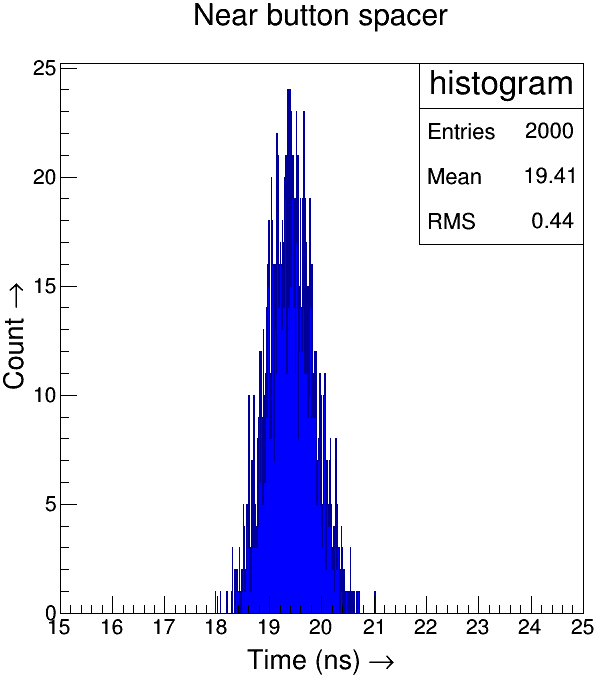}
  \label{fig:timingHisto_nearBSpacer}
  }
 \caption{Timing histograms for \subref{fig:timingHisto_regular} regular region, \subref{fig:timingHisto_nearESpacer}
  1 mm from edge spacer and \subref{fig:timingHisto_nearBSpacer} 200 $\mu$m from the corner 
  of pedestal part of the button spacer.}
\label{fig:timingHisto}
\end{figure}
 The histograms of electron drift times for different regions are shown 
in figure \ref{fig:timingHisto}. The mean of each distribution gives the average time taken by the electrons to reach the plate on which
voltage of positive polarity has been applied. From the three histograms it can be seen that the 
electrons at 1 mm away from the edge spacer (E$_{z}$ = 36.02 kV/cm) take about 3 ns longer time 
whereas, the electrons at 200 $\mu$m away from the corner of the pedestal part of the button spacer 
(E$_{z}$ = 35.72 kV/cm) take about 2 ns more in comparison to those drifting in the regular region.
\\To see the effect of SF$_{6}$ on the timing properties, the same calculations have been performed 
for the gas mixtures R-134A : Isobutane : SF$_{6}$ = 95.0 : 4.5 (4.8, 4.9) : 0.5 (0.2, 0.1). The 
variation of the two timing parameters with the applied voltage and the corresponding value of 
E$_{z}$ is shown in figure \ref{fig:t_sigma_vs_HV_manyGas}.
A higher value of both average signal arrival time and time resolution has been found for higher 
amounts of SF$_{6}$ in the gas mixture which is at par with our earlier work \cite{paper-salim}.

  \section{Conclusion}
\label{sec:conclusion}
The reduction of signal amplitude as one approaches the edge spacer can be attributed to  
the affected field configuration of those regions and will lead to loss of response decreasing the 
effective active area of a RPC.
\\
The trend of variation in signal arrival time and time resolution with the applied voltage in simulation 
agrees with that obtained in experiment, although they differ quantitatively. Nevertheless, the 
present simulation produces time resolution values close to the analytical estimates \cite{paper-Riegler_timing}. 
The simulation will be improved in the future by considering the real-life factors of an experiment, 
to compare with the measurement. This will include the effect of movement of ions 
and electronic impedance on the signal shape. We also plan to take the finite bulk resistivity 
of the bakelite plates into account instead of treating them as perfect dielectrics.
\\
Preliminary calculations have shown that electrons at the critical regions like near an edge or button 
spacer, takes more time compared to a region away from any imperfections. This may 
affect the timing response in those critical regions. The effect of these regions on the RPC time 
resolution following the present method of calculation by considering a current threshold will be 
investigated in the future.
\\
The average signal arrival time has been found to increase with the amount of SF$_{6}$ present 
in the gas mixture. The present calculations have shown a deterioration in time resolution with 
the increase in SF$_{6}$ percentage.
  \acknowledgments

We would like to acknowledge the fruitful discussions with the INO collaboration and 
thank Purba Bhattacharya for her suggestions with the numerical calculations and 
Meghna K.K. for her help in the experimental measurements. We also thank the reviewer of this 
paper and the Editor for their constructive suggestions and help in improving the quality 
of the paper.

\end{document}